\documentclass[pre,twocolumn,showkeys,showpacs,amsfonts,amssymb,amsmath,letterpaper]{revtex4}

\usepackage{graphicx}

\usepackage{bm}

\renewcommand{\vec}[1]{\bm{#1}}

\newcommand{\D}[2]{\frac{\partial #1}{\partial #2}}

\newcommand{\RD}[2]{\frac{\mathcal{D} #1}{\mathcal{D} #2}}

\begin{document}

\title{Dynamics of shear-transformation zones in amorphous plasticity:\\
large-scale deformation in a two-dimensional geometry}

\author{L. O. Eastgate}
\affiliation{Laboratory of Atomic and Solid State Physics, Cornell University,
Ithaca, New York 14853}
\altaffiliation{Current address: Fredensborgveien 41, NO-0177 Oslo, Norway.}
\date{September 2005}

\begin{abstract}
A two-dimensional version of the shear-transformation-zone (STZ) theory by Falk and
Langer is explored numerically. Two different geometries are used to simulate uniaxial
tension experiments where materials are subjected to constant strain rates. In the
first setup, a rectangular specimen is given an imperceptible indentation, allowing it
to neck at the center. The dynamics are explored systematically by varying both the
straining capability of the STZs ($\epsilon_0$) and the external strain rate. Higher
values of $\epsilon_0$ increase the plastic flow and result in sharper
necks. Decreased values of the external strain rate result in the formation of narrow
shear bands, consistent with the absence of thermal relaxation mechanisms in the
model.  In the second configuration, the equivalent of pre-annealed materials are
explored. Here, the sample is initially square and the edges are made rough in order
to encourage the formation of shear bands. With respect to $\epsilon_0$ and the
external strain rate, the results show trends similar to those in the necking
simulations. The pre-annealing was modeled by using a low initial density of
STZs. This had most effect when $\epsilon_0$ was large, contributing to the
localization of the strain and making the material appear more brittle.
\end{abstract}





\maketitle

\section{Introduction}
\label{sec:introduction}
The goal of the continuum shear-transformation-zone (STZ) theory is to describe
plastic deformation in amorphous solids on a mesoscopic scale, averaging out some of
the microscopic, discrete details.  The plastic deformation is described in terms of
flow rates, similar in approach to that of the Navier-Stokes model
\cite{LANDAU_FLUIDMECHANICS} although the scalar pressure has been replaced with a
stress tensor.

The STZ theory was constructed by Falk and Langer \cite{FALK-THESIS,
FALK_1998_PRE_57_6_7192, FALK_1999_PRB_60_10_7062, FALK_2000_MB_25_5_40,
LANGER_1999_PRE_60_6_6978, LANGER_2000_PRE_62_1_1351, LOBKOVSKY_1998_PRE_58_2_1568,
LANGER_2001_PRE_6401_1_011504},
and originated with the assumption that plastic deformation is limited to, and defined
as, the non-affine transformation of particles in localized areas or zones. This was
based on similar ideas made by Argon, Spaepen, and others who described creep in
metallic glasses in terms of local, molecular transitions or rearrangements
\cite{ARGON_1979_MSE_39_1_101, SPAEPEN_1977_AM_25_4_407, SPAEPEN_LESHOUCHES,
ARGON_1979_AM_27_1_47,
ARGON_1983_AM_31_4_499, KHONIK_1994_JNS_170_3_270}. This, in
turn, grew out of theories by Turnbull, Cohen, and others, who suggested that the
observed behavior in the amorphous metals could be described by linking the transition
rates to local free-volume fluctuations \cite{SPAEPEN_LESHOUCHES,
COHEN_1959_JCP_31_5_1164, TURNBULL_1961_JCP_34_1_120, TURNBULL_1970_JCP_52_6_3038}.

This paper, continuing the work reported in \cite{EASTGATE_2003_PRL_90_4_045506},
presents a two-dimensional model that describes elastic and plastic
dynamics in an amorphous solid; the STZ theory supplies the model with a plastic-flow
description. The main contribution of this paper is the exploration of the STZ theory
in a spatially extended system, focusing on features introduced by having
an inhomogeneous geometry.
Section~\ref{sec:tensorial_stz}
presents a tensorial version of the STZ theory that was developed in
\cite{LANCE-THESIS} and based on both the scalar model developed earlier by Falk and
Langer, as well as some of Falk's initial
ideas of how to write the theory in a two-dimensional setting \cite{FALK-THESIS}.

The remaining part, which forms the core of the paper, reviews two sets of
numerical uniaxial-tension constant-strain-rate simulations of the two-dimensional,
tensorial STZ model. First, Section~\ref{sec:numerics} gives a brief discussion of the
implementation. Then the first set is presented in Section~\ref{sec:necking},
exploring how the plastic flow affected the dynamics during necking. The second set is
described in Section~\ref{sec:annealed}, relating the pre-annealing of amorphous
materials to increased strain localization and brittle behavior.
The paper is concluded in Section~\ref{sec:conclusion}.


\section{The Tensorial STZ Model}
\label{sec:tensorial_stz}
The numerical simulations described in this paper uses the so-called quasi-linear
tensorial STZ model in two dimensions, given below. When combined with boundary
conditions, it can be used to describe both the elastic and plastic dynamics of an
amorphous solid in a spatially extended geometry. The equations are
\begin{subequations}
\label{eq:complete_eqn_mot}
\begin{align}
\label{eq:complete_pdot}
\left(\frac{1-\nu_2}{1+\nu_2}\right)\frac{1}{2\mu}\frac{Dp}{Dt}
&= -\frac{1}{2}\vec{\nabla}\cdot\vec{v}\,,\\
\label{eq:complete_sdot}
\frac{1}{2\mu}\RD{s_{ij}}{t} &= \tilde{D}^\text{tot}_{ij}-D^\text{pl}_{ij}\,,\\
\label{eq:complete_vdot}
\rho\frac{Dv_i}{Dt} &=\frac{\partial s_{ji}}{\partial x_j}
 -\frac{\partial p}{\partial x_i}\,,\\
\label{eq:complete_Lambdadot}
\frac{D\Lambda}{Dt} &= \Gamma(1-\Lambda)\,,\\
\label{eq:complete_Deltadot}
\RD{\Delta_{ij}}{t} &= \frac{1}{\epsilon_0}D^\text{pl}_{ij}-\Gamma\Delta_{ij}\,,
\end{align}
with
\begin{align}
\tilde{D}^\text{tot}_{ij} &= \frac{1}{2}\left(\D{v_i}{x_j}+\D{v_j}{x_i}\right)
 -\frac{1}{2}\vec{\nabla}\cdot\vec{v}\,,\\
\frac{1}{\epsilon_0}D^\text{pl}_{ij}
&=
\left(
\Lambda s_{ij}-\Delta_{ij}\right)\,,
\end{align}
and
\begin{equation}
\label{eq:Gamma}
\Gamma =
\frac{2\Lambda
\left(\frac{1}{\epsilon_0}\bar{D}^\text{pl}\right)^2}
{
(1+\Lambda)(\Lambda^2-\bar{\Delta}^2)}\,.
\end{equation}
\end{subequations}

The variables here are the pressure $p$, the deviatoric stress $s_{ij}$, the velocity
$v_i$, the relative density of STZs $\Lambda$, and the STZ alignment
$\Delta_{ij}$. $\tilde{D}_{ij}^\text{tot}$ and $D_{ij}^\text{pl}$ are total and
plastic rates of deformation, respectively. The tilde on the former indicates that
only the deviatoric part of the tensor is being used (that is, the trace has been
subtracted from it). $\Gamma$ is proportional to both the rate at which STZs are
created and annihilated, as well as to the rate at which energy is dissipated through
plastic deformations. The invariant of the rate of plastic deformation is defined as
$\bar{D}^\text{pl}=\sqrt{D^\text{pl}_{kl}D^\text{pl}_{kl}/2}$. The parameters in the
equations, which remain constant throughout a simulation, are the shear modulus $\mu$,
the two-dimensional Poisson ratio $\nu_2$, the mass density $\rho$, and the straining
capability of the STZs $\epsilon_0$. The temporal derivatives $D/Dt$ and
$\mathcal{D}/\mathcal{D}t$ both include an advective term, and the latter also takes
rotation into account. The complete derivation of these exact equations can be found
in \cite{LANCE-THESIS}; see also \cite{LANGER_2003_PRE_68_6_061507,
FALK_2004_PRE_70_1_011507, LANGER_2004_PRE_70_4_041502, PECHENIK_PRE_72_2_021507}.

Concerning expressions for the stored energy, it was found in an earlier publication
\cite{LANGER_2003_PRE_68_6_061507} that the plastic dissipation is given by
\footnote{Note that 
\cite{LANGER_2003_PRE_68_6_061507} has some slight differences in definitions:
$Q_\text{Pechenik}=Q_\text{Lance}$, but $\Gamma_\text{Pechenik} =
2\Gamma_\text{Lance}$ and
$\psi^\text{pl}_\text{Pechenik}=\psi^\text{pl}_\text{Lance}$. Since the $\Gamma$s are
different, the equations of motion are slightly different as well, including the
limiting case for $\dot{\Delta}$ as $\Lambda\to 1$.}
\begin{equation}
\label{eq:Q}
Q = 
2\epsilon_0\Lambda\Gamma\,,\\
\end{equation}
and the plastic energy density
\begin{equation}
\label{eq:stored_plastic}
\psi^\text{pl} = \epsilon_0 
\left(\frac{\Lambda^2+\bar{\Delta}^2}{\Lambda}\right)\,.
\end{equation}
In addition, the elastic and kinetic energy densities are given by
\begin{align}
\label{eq:stored_elastic}
\psi^\text{el} &= \frac{\sigma_{ij}\varepsilon_{ij}}{2}
=\frac{1}{2\mu}\left[\left(\frac{1-\nu_2}{1+\nu_2}\right)p^2 + \bar{s}^2\right]\,,\\
\label{eq:kinetic_energy_density}
\psi^\text{kin} &= \frac{1}{2}\rho\vec{v}^2
= \frac{1}{2}\rho v_x^2 + \frac{1}{2}\rho v_y^2\,.
\end{align}

One possible interpretation of how plastic deformation \emph{might} be described by
the STZ model, involves the flipping of STZs. A deformation caused by some STZs
flipping, will encourage the internal stress distribution to change, which in turn
will flip even more STZs \emph{as well as} create new ones and annihilate existing
ones. It takes energy to flip an STZ. The work done on it could even be considered
reversible, if one could make the STZ flip back again. Then again, the STZ could make
it back to its original position without releasing any energy by being annihilated and
recreated, corresponding to permanent deformation. Since, theoretically, one could
regain the energy in the flipped STZ if one could make all the STZ flip back again
without any of them being annihilated, a reasonable interpretation of \emph{stored
plastic energy} $\psi^\text{pl}$ could be the energy stored in the flipped STZs. In
practice, though, plastic deformation is an irreversible process; some of the stored
plastic energy would be lost if the STZs were to flip back, since this flipping would
cause further deformations and thus more creation and annihilation of STZs. The
plastic dissipation $Q$ can be interpreted as the energy lost when an STZ is
annihilated and recreated.

Simplified, zero-dimensional models using a non-tensorial version of the STZ equations
have been explored in earlier publications \cite{FALK_1998_PRE_57_6_7192,
FALK_1999_PRB_60_10_7062, LOBKOVSKY_1998_PRE_58_2_1568, LANGER_1999_PRE_60_6_6978,
LANGER_2000_PRE_62_1_1351}.  The current manuscript concentrates on the effects of a
two-dimensional geometry.


\section{Numerical Setup}
\label{sec:numerics}
In order to explore the tensorial version of the STZ theory given in
Eqs.~\eqref{eq:complete_eqn_mot}, a couple of 
two-dimensional geometrical configurations were implemented and simulated
numerically. For this purpose, a C++-program using finite-difference algorithms on a
regular grid with a second-order explicit time-stepping scheme to integrate the
equations was written from scratch. This section will give a brief overview over the
numerical implementation of the model, including the mapping of the variables onto a
unit square, numerical algorithms and approximations, and boundary conditions.

\begin{figure}
\begin{center}
      \includegraphics[width=\columnwidth,clip]{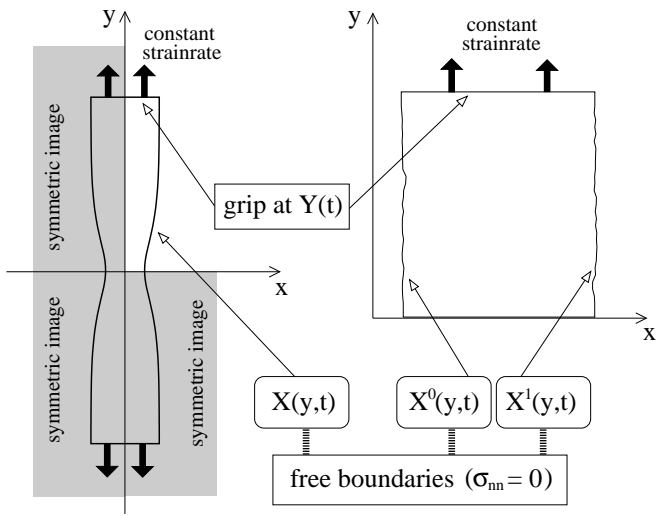}
      \caption
      {The geometrical setup of the two-dimensional simulations. A constant
      strain rate was applied to the rigid ends at $Y(t)$ by adjusting the
      velocity. The free boundaries at $X(y,t)$, $X^0(y,t)$, and $X^1(y,t)$ had no
      normal stresses, that is, $\sigma_{nn}=0$. In the necking simulations on the
      left, the material was assumed to be symmetric across the $x$ and $y$ axes; thus
      only the upper right-hand corner of the material was actually simulated. In the
      simulations on the right, the lower edge was held fixed.}
      \label{fig:necking_and_free_setup}
\end{center}
\end{figure}

The two-dimensional STZ theory was implemented numerically in order to simulate
uniaxial tension experiments where material specimens were strained at constant
rates. The velocity was controlled along the top and bottom edges (the ``grips''),
while the left and right boundaries were assumed to have no normal stresses and were
allowed to deform; see Fig.~\ref{fig:necking_and_free_setup}.

Two sets of simulations were performed; in the first set the material was made to
neck (Section~\ref{sec:necking}), while in the second set the samples were
given rough boundaries to encourage the formation of shear bands
(Section~\ref{sec:annealed}). In the necking 
simulations, a small indentation was applied at the center of the material to break the
symmetry and to encourage the neck to form in the middle. In addition, the material
was assumed to be symmetric across both the $x$- and $y$-axis,
so only a quarter of the system was modeled numerically. 
In the simulations with rough boundaries, the left and right
edges were made stress-free and independent of each other, and no
symmetry assumptions were used. The bottom of the material was held fixed while the
top was subjected to a constant strain rate.

\begin{figure}
\begin{center}
      \includegraphics[width=\columnwidth,clip]{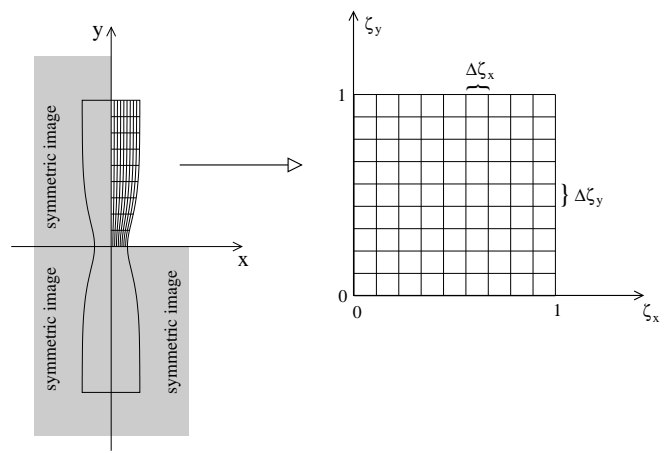}
      \caption
      {For the necking simulations, the deformed area $x\in[0,X(y,t)]$,
      $y\in[0,Y(t)]$ was mapped onto the unit square $\zeta_x\in[0,1]$,
      $\zeta_y\in[0,1]$. Similarly, the non-symmetric simulations (not shown here)
      mapped the area 
      $x\in[X^0(y,t),X^1(y,t)]$, $y\in[0,Y(t)]$ onto the same unit square.}
      \label{fig:discrete}
\end{center}
\end{figure}

One aim when introducing the two-dimensional model was to find out how it differed
from 
the zero-dimensional description, in particular with respect to geometrical
inhomogeneities. After deciding to investigate constant strain rate simulations, it
seemed sufficient to only allow the sides to deform, while keeping the grips
straight. It was assumed that the free boundaries could be described by functions that
were single valued: $X(y,t)$ for the necking simulations, or 
$X^0(y,t)$ and $X^1(y,t)$ in the case of the model without symmetry assumptions.
The position of the grips was described by a function only dependent on
time, $Y(t)$. With these assumptions, the material was easily mapped onto a unit
square with coordinates $\zeta_x\in[0,1]$ and $\zeta_y\in[0,1]$. For the necking
simulations with the axis symmetries, the mapping was
\begin{align}
\label{eq:transform_1}
\zeta_x   &= \frac{x}{X(y,t)}\,, &
\zeta_y   &= \frac{y}{Y(t)}\,,
\end{align}
while the simulations with the free boundaries needed the slightly more general
transformation
\begin{align}
\label{eq:transform_2}
\zeta_x   &\equiv \frac{x-X^0(y,t)}{X^1(y,t)-X^0(y,t)}\,, &
\zeta_y   &\equiv \frac{y}{Y(t)}\,.
\end{align}
See Fig.~\ref{fig:discrete} for an illustration.

The advantage of mapping the coordinates using Eqs.~\eqref{eq:transform_1} and
\eqref{eq:transform_2} was that the calculations could be done on a regular grid where
the values of the fields were discretized into equally spaced nodes. The disadvantage
was that the operators, like the first-order derivatives, were more complicated to
calculate.

All the simulations were done at constant strain rate. This was enforced by controlling
the $y$-component of the velocity, $v_y$, at the grips (the top and bottom boundaries
in Fig.~\ref{fig:necking_and_free_setup}). For example, in the necking simulations the
grip velocity was given by
\begin{equation}
v_y[x,y=\pm Y(t),t] = \pm Y(t) D^\text{tot}\,,
\end{equation}
where $D^\text{tot}$ was the constant strain rate. Since the velocity was applied to
all the nodes along the top and bottom, these edges remained straight. The position of
the edge as a function of time, using that $v_y[x,y=Y(t),t] = \partial_tY(t)$, was
therefore
\begin{equation}
Y(t) = Y(0)\exp\left(t D^\text{tot}\right)\,.
\end{equation}
In addition to controlling $v_y$, the variables $\tau$ and $\Delta_\tau$ were kept at
zero along the grips, effectively removing any shear there.

At the free boundaries, which comprised $X(y,t)$ in the necking simulations and
$X^0(y,t)$ and $X^1(y,t)$ in the others, the normal stress $\sigma_{nn}$ and shear
stress $\sigma_{nt}$ were set to zero, while the tangential stress $\sigma_{tt}$ was
left untouched.

The simulations were carried out on a uniform rectangular grid using
fi\-nite-diff\-erence
approximations for the spatial derivatives.
The fields were integrated forward in time using the mid-point method, also known as
the second-order Runga-Kutta scheme, as well as an adaptive time-step technique called
step-doubling~\cite{NR}.


When implementing the STZ equations of motion, a numerical viscosity
$\eta\vec{\nabla}^2\vec{v}(\zeta)$ was added to the velocity variables on the mapped
grid. This was needed because the first-order derivatives were discretized using a
central-difference algorithm; with no higher-order spatial derivatives, the
grid-points had a tendency to ``decouple'' into two sets of nodes like the black and
white squares on a chess board.  Note that since the Laplacian was applied to the
velocities on the mapped grid (that is, on the unit square) and not to the usual
velocities, the diffusion term would only be a ``proper'' viscosity where the material
was undeformed.

Since the added viscosity was only meant as a numerical tool, it was important to
monitor any changes it made to the physical results. The influence of the viscosity
was minimized by performing successive runs with smaller and smaller values of $\eta$.
If the numerical viscosity was chosen too high, the results would look nice and
smooth, but when compared to simulations with lower values of $\eta$ it became clear
that the added viscosity was smearing out sharp features such as shear bands. On the
other hand, when too small values of $\eta$ were used, large gradients caused by the
``decoupling'' would make the simulations more unstable. This was particularly
pronounced at the border (where values were calculated through extrapolation) and in
larger grids.

The simulations were most sensitive to the value of $\eta$ at low strain rates,
basically because there was more time (per strain) to dampen out the velocities.  When
running as slow as $D^\text{tot}=10^{-5}$ on the larger (more unstable) grids used in
Section~\ref{sec:annealed}, there were no values of $\eta$ that gave satisfactory
results; even the smallest stable value of $\eta$ would be too large, resulting in
unacceptable amounts of artificial dissipation and influencing the behavior too much.
Even for the smaller grids in the necking simulations of
Section~\ref{sec:necking}, the slowest strain rate $D^\text{tot}=10^{-5}$ was hard to
accommodate. In the end, it was found that $\eta=0.02$ (which in fact turned out to be
a good choice for all the simulations) enabled the slowest simulation to reach $5\%$
strain while having only a small impact on the physical results. The
influence due to the exact choice of $\eta$ for higher strain rates was usually
imperceptible.


\section{Necking}
\label{sec:necking}
Irregular geometry in a two-dimensional, externally loaded sample of material can
cause stress 
localizations and shear banding. Compared to a zero-dimensional model, two
spatial dimensions add elastic deformation and inertia. As 
long as the elastic strains are small, the speed of sound high, and the rate of
total deformation low, the difference between a zero-dimensional and a uniform
two-dimensional system is minimal.
Generally speaking, metallic glasses have two modes of deformation: homogeneous and
inhomogeneous \cite{SPAEPEN_BOOK, 
ARGON_1982_JPCS_43_10_945, ARGON_1983_AM_31_4_499, TAUB_1981_JMS_16_11_3087,
LU_2003_AM_51_12_3429}. A transition from the former to the latter will occur when a
material is strained and the stress localizations from the irregular geometry start
causing shear bands. The simulations described in this
and the next section illustrates this transition. Some earlier work can be found
in~\cite{EASTGATE_2003_PRL_90_4_045506}.

\subsection{Simulations}
\label{sec:simulations}
One way of exploring the effect of a non-trivial geometry is to look at the dynamics
of a material while it is necking. A series of simulations were run where a
rectangular $2\times 8$ piece of material was elongated in the $y$-direction at a
constant strain rate.\footnote{All the values, including lengths, have been normalized
and made unitless. To recover all the physical values with units in a specific system,
three material properties are needed, for example the density $\rho$, the shear
modulus $\mu$, and the length of the material sample.}  The material was slightly
indented in the middle, where the width was reduced by $1\%$. Specifically, the
right-hand boundary was given by $X(y,t=0) = 1 - \delta(y)$, where
$\delta(y)=0.01\exp[-\ln 2 (y/0.1)^2]$. This perturbation to the geometry was added to
break the symmetry and to help trigger any potential instabilities. The material (or
rather, a quarter of it) was mapped onto an $11\times 41$ regular grid, using the
transformation of variables described in Section~\ref{sec:numerics}. The size of the
grid was verified to be large enough to support the desired accuracy, as a comparison
with some test runs on a larger $21\times 81$ grid gave almost identical results. The
shear modulus was set to $\mu=100$ and the density $\rho=1$; since the speed of sound
is proportional to $\sqrt{\mu/\rho}$, a lower value of $\mu$ would have jeopardized
the assumption that the elastic deformations were instantaneous. Conversely, an
increased value of $\mu$ would demand notably more resources computationally. All the
necking simulations had $\Lambda_0=1$, and the numerical viscosity was chosen to be
$\eta=0.02$, based on the discussion in Section~\ref{sec:numerics}.

\begin{figure}
\begin{center}
      \includegraphics[height=0.4\textheight,clip]
      {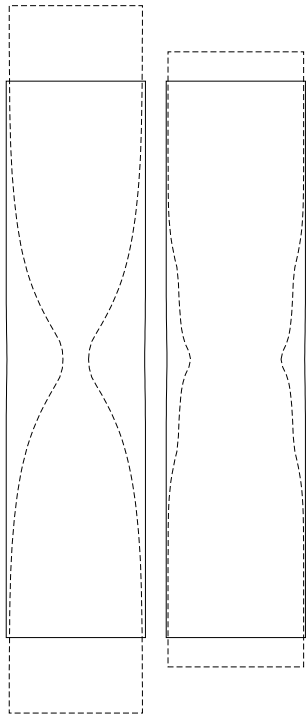}
      \caption{Examples of the geometry in two numerical
      experiments. The solid outlines are the initial configurations --- the initial
      indentations are hardly visible. The dashed outlines represent the geometry
      after the left and right samples were strained $24\%$ and $10\%$,
      respectively. The strain rate in the left simulation was $D^\text{tot}=10^{-3}$,
      ten times higher than on the right. In both simulations $\mu=100$, $\nu_2=0.5$,
      $\epsilon_0=0.03$, $\Lambda_0=1$, and the initial size of the samples were
      $2\times 8$.}
      \label{fig:two_examples}
\end{center}
\end{figure}

Fig.~\ref{fig:two_examples} shows the outlines of material samples in two separate
simulations.
The solid outlines were the initial configurations; notice that the
indentations are hardly visible. The dashed outlines represent the configurations
after the systems had been strained. The simulation on the left was strained an order
of magnitude faster than the one on the right. While the former had a smooth neck, the
latter had a more irregular boundary due to shear bands.

\begin{figure}
\begin{center}
  \begin{tabular}{cc}
      \includegraphics[bb = 213 112 247 246, height=0.4\textheight,clip]
        {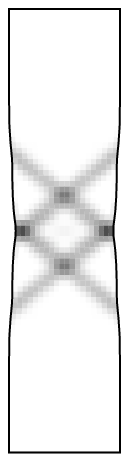}
      &\includegraphics[height=0.4\textheight,clip]
        {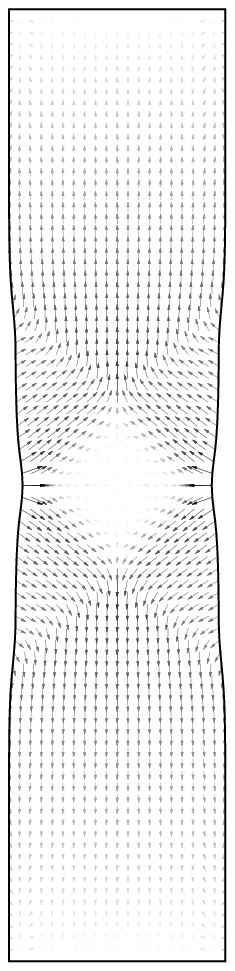}
  \end{tabular}
      \caption
      {The plots in this figure are taken from the same simulation as the outline on
      the right-hand side of Fig.~\ref{fig:two_examples}, but this time at roughly
      $6\%$ strain. The density plot on the left shows the plastic dissipation $Q$ as
      given by Eq.~\eqref{eq:Q}. The figure
      on the right displays the velocity field after the uniform part
      $\vec{v}_\text{uniform}=(0,y D^\text{tot})$ had been subtracted away.
      ($D^\text{tot}=10^{-4}$, $\epsilon_0=0.03$, $\Lambda_0=1$)}
      \label{fig:Q_and_v}
\end{center}
\end{figure}

In general, shear bands would appear more readily when the material was strained at a
lower rate. The left part of Fig.~\ref{fig:Q_and_v} shows the plastic dissipation $Q$
as given by Eq.~\eqref{eq:Q} in a simulation with $D^\text{tot}=10^{-4}$,
$\epsilon_0=0.03$, and $\Lambda_0=1$ after it had strained roughly $6\%$.
The right-hand side displays, for the same simulation, the velocity field with the
uniform part
$\vec{v}_\text{uniform}=(0,y D^\text{tot})$ subtracted away. These figures
show that the material no longer was deforming uniformly. Stress concentrations
radiating out from each notch at roughly $45^\circ$ increased the plastic dissipation,
and thus the plastic deformation, along these shear bands. As can
be seen in the velocity plot, the whole central piece of the neck became
narrower, and this is what caused
the irregular boundaries seen on the right-hand side of Fig.~\ref{fig:two_examples}.

\begin{figure}
\begin{center}
      \fbox{
      \begin{tabular}{l}
        \includegraphics[width=\columnwidth,clip]{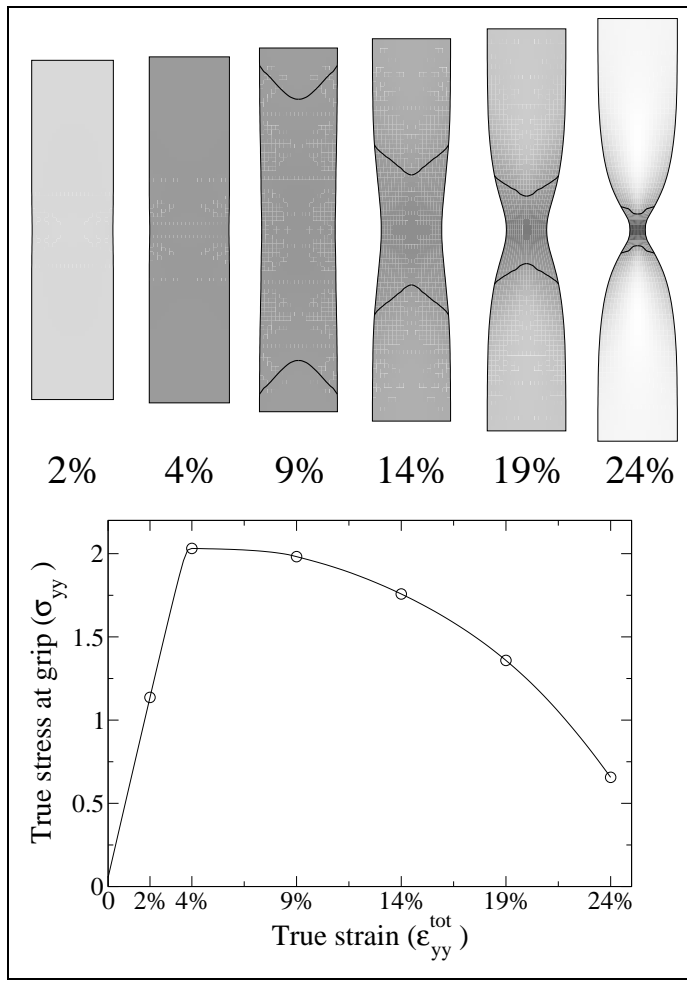}\\
        \ \\
        \includegraphics[width=0.93\columnwidth,clip]{figs/ss_true_grip_dtot0-001_Lambda1_eps0-03}
      \end{tabular}}
      \caption
      {The graph shows the true stress $\sigma_{yy}$ at the grips as a
      function of the total strain $\varepsilon_{yy}^\text{tot}$. Snapshots of the
      system were
      taken at six different strains, and the interior was shaded according to the
      value of the maximum shear stress $\bar{s}=\sqrt{s^2+\tau^2}$. The black lines
      in the four last snapshots show where $\bar{s}=1$, suggesting that the area
      between the lines ($\bar{s}>1)$ was flowing plastically while the areas at the
      ends ($\bar{s}<1$) were jammed. ($D^\text{tot}=10^{-3}$,
      $\epsilon_0=0.03$, $\Lambda_0=1$)}
      \label{fig:sbar}
\end{center}
\end{figure}

In a uniform STZ material, the rate of deformation $D^\text{pl}$
increases drastically when the deviatoric stress $|s|$ becomes larger than unity
\cite{LANCE-THESIS, EASTGATE_2003_PRL_90_4_045506, LANGER_2003_PRE_68_6_061507}. In
steady state, any area with $|s|>1$ can be considered to be flowing plastically, while
regions with $|s|<1$ are jammed. This applies to the tensorial version of the theory,
too, when measuring the deviatoric stress by the invariant $\bar{s}=\sqrt{s^2+\tau^2}$,
also known as the maximum shear stress. Fig.~\ref{fig:sbar} shows a series of
snapshots from a necking simulation where the interior has been shaded according to
the values of $\bar{s}$; the darker the shade, the higher the value (notice that the
strain is plotted against the total stress $\sigma_{yy}$ at the grip, and since
$\sigma_{xx}=\sigma_{xy}=0$ there, one has that $\sigma_{yy}=2\bar{s}$).
At low strains the stress was uniform, but as the material necked the stress
concentrated in the center. The black lines that appear in the last four snapshots
mark the boundary between $\bar{s}<1$ and $\bar{s}>1$. The steady-state solutions
suggest that the area between the lines were flowing plastically, while the regions on
either end, outside the lines, were more or less jammed. As the neck grew more
pronounced, the plastically flowing area would shrink, making a smaller and smaller
area responsible for accommodating the ever increasing global strain. In addition, the
jammed areas would relax (notice how the ends grew lighter in the final snapshots) and
contract along the strained $y$-axis while releasing stored
energy. This decrease in strain at the ends had to be compensated by increasing the
strain even further in the middle.

The graph below the snapshots plots the true stress $\sigma_{yy}$ at the grips as a
function of the total true strain $\varepsilon^\text{tot}_{yy}$. The true stress (at
the grip) is defined as the 
force divided by the current width of the material, both measured at the grip. The
true, or logarithmic, total strain is defined as
$\varepsilon^\text{tot}_{yy}=\ln[L_y(t)/L_y(0)]$, where $L_y(t)$ is the length of the
material at time $t$.  The snapshots were marked on the graph as circles. The first
snapshot was taken while the material was in the ``elastic phase'', while the second
was taken just as the material was about to yield. This explains why the two first
snapshots seem so uniform. Note that since the plotted stress was measured at the
grips, it corresponds to the shading at the ends of the material.

\begin{figure}
\begin{center}
      \includegraphics[width=\columnwidth,clip]{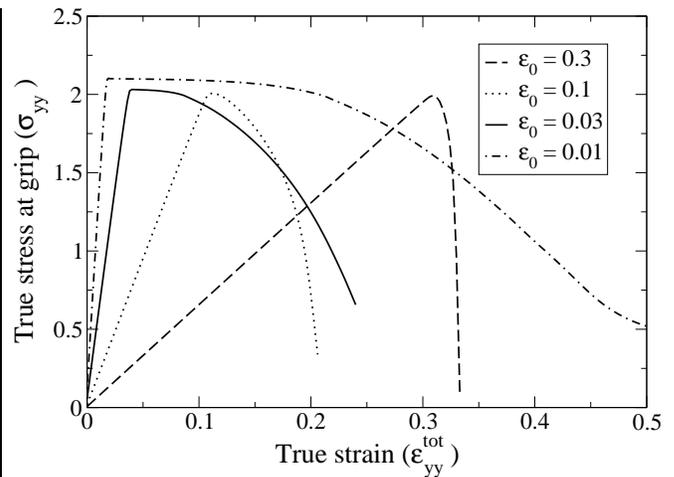}
      \caption
      {The true stress $\sigma_{yy}$ at the grip plotted against the true
      total strain $\varepsilon^\text{tot}_{yy}$ for selected values of
      $\epsilon_0$. As $\epsilon_0$ was increased, it took longer (in terms of
      strain) for the material to reach the yield stress, since less of the
      deformation was stored as elastic
      energy (and more as plastic). Once the yield stress was reached, a higher value
      of $\epsilon_0$ allowed 
      for more plastic deformation and thus a quicker relaxation of the stress.
      ($D^\text{tot}=10^{-3}$, $\mu=100$, $\nu_2=0.5$, $\Lambda\equiv 1$)}
      \label{fig:ss_true_dtot0.001}
\end{center}
\end{figure}

\begin{figure}
\begin{center}
      \includegraphics[width=\columnwidth,clip]{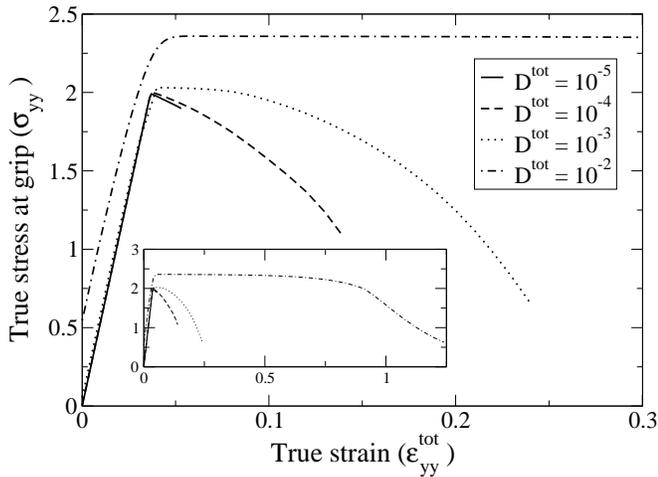}
      \caption
      {The true stress $\sigma_{yy}$ at the grip plotted against the true
      total strain $\varepsilon^\text{tot}_{yy}$ for selected values of
      $D^\text{tot}$. The value
      of $D^\text{tot}$ had little effect on the ``rate'', in terms of strain, at
      which the stress reached yield; that is to say, all the simulations in this
      graph yielded approximately at the same strain (the reason the
      $D^\text{tot}=10^{-2}$ curve starts at a higher stress is explained in the
      text). After yield, though, the
      high strain rate simulations were strained much further before they relaxed (see
      framed plot).
      ($\epsilon_0=0.03$, $\mu=100$, $\nu_2=0.5$, $\Lambda\equiv 1$)}
      \label{fig:ss_true_eps0.03}
\end{center}
\end{figure}

\begin{figure}
\begin{center}
      \includegraphics[width=\columnwidth,clip]{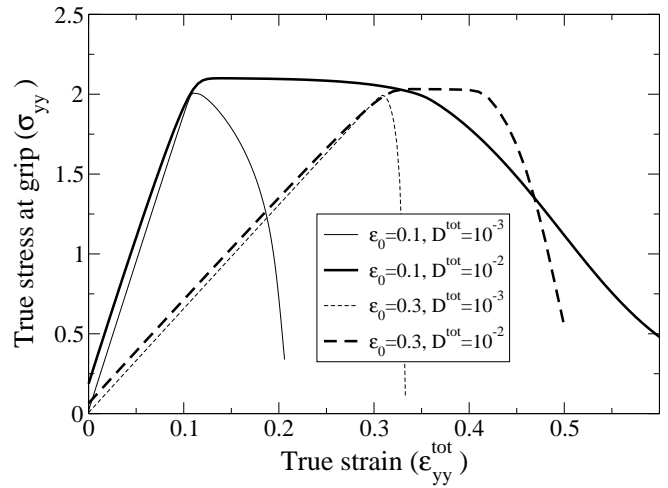}
      \caption
      {The true stress $\sigma_{yy}$ at the grip plotted against the true total strain
      $\varepsilon^\text{tot}_{yy}$. Again, a higher
      $\epsilon_0$ caused a slower rise to yield but a faster relaxation after. The
      strain rate did not 
      affect the rise to yield, but a faster rate resulted in longer plateaus of
      ``uniform steady-state plastic deformation'' before the indented geometry caused
      a necking instability. ($\mu=100$, $\nu_2=0.5$, $\Lambda\equiv 1$)}
      \label{fig:ss_true_Lambda1}
\end{center}
\end{figure}

To illustrate how the parameters affect the material behavior,
Figs.~\ref{fig:ss_true_dtot0.001}, \ref{fig:ss_true_eps0.03}, and
\ref{fig:ss_true_Lambda1} show further plots where the true stress 
$\sigma_{yy}$ at the grip was plotted against the true total strain
$\varepsilon^\text{tot}_{yy}$, comparing curves for various values of $\epsilon_0$ and
$D^\text{tot}$.
In Fig.~\ref{fig:ss_true_dtot0.001} all the simulations were strained at a rate of
$D^\text{tot}=10^{-3}$ while $\epsilon_0$, the amount of strain caused by flipping
STZs, was varied. In simulations with a higher $\epsilon_0$, the material needed to be
strained further before reaching the yield stress. This is because more of the work
done on the system was stored as plastic energy; it was only the elastic deformations
that contributed to the rising stress. In fact, the slope of the stress-strain
curves in the ``elastic'' regime, which corresponds to the \emph{effective} shear
modulus, has been calculated in \cite{FALK_2004_PRE_70_1_011507}.
After reaching the yield stress, the systems
with a higher $\epsilon_0$ would see a faster relaxation of the stress. This is
probably because a high $\epsilon_0$ allowed for a higher plastic rate of deformation
at the neck, where the elastic stored energy was released through plastic
dissipation. Interestingly enough, a high $\epsilon_0$ meant that most of the stored
energy was plastic; that is, the energy was stored in flipped STZs rather than elastic
strain (the amount of stored plastic energy cannot easily be deduced from the
stress-strain curves).

The effect of varying the total strain rate is shown in
Fig.~\ref{fig:ss_true_eps0.03}. The strain rate had little effect
on the strain at which the yield stress was reached (although
it did influence the initial stress in these simulations; see below).

After the system reached yield stress, the simulations with high
strain rates took a lot longer (in terms of strain) before they would neck. The higher
strain rate meant a steady-state stress even further above unity (the ``yield
stress''), which in turn meant
that small geometric inhomogeneities would not be able to separate the stress into
regions of $\bar{s}>1$ and $\bar{s}<1$. For the low strain rate simulations, the stress
localization was more pronounced, and for the lowest rate, the shear bands were so
sharp that the numerics was not able to strain the material beyond $5\%$.

Fig.~\ref{fig:ss_true_Lambda1} compares stress-strain curves of
simulations where both $\epsilon_0$ and $D^\text{tot}$ were varied. These curves show
clearly that the strain rate primarily changed the time between yield and necking,
while $\epsilon_0$ controlled the dynamics of both the stress increase before yield
and the relaxation after.

On a side note, all the simulations were started at the given (or final) strain rate
rather than being ramped up from zero. This was especially important for the high
strain rate simulations. If the material had been started from rest, the details of
how the strain rate was ramped up would greatly affect the outcome of the
simulation. If ramped up too slowly, the material would yield before the final strain
rate could be achieved. If the strain rate was turned up too quickly, the speed of
sound would no longer be negligible and inhomogeneities would form at the grips due to
large stress buildups.  To minimize transient effects from the non-zero initial strain
rate (such as standing elastic waves in the material), initial values for the other
variables in the simulation were calculated in order to start the system as close as
possible to a steady state. This is why the curve with a strain rate of
$D^\text{tot}=10^{-2}$ in Fig.~\ref{fig:ss_true_eps0.03} starts with a somewhat higher
stress. This is not a numerical effect; it would be true for real experiments as
well. See Fig.~\ref{fig:phase_diagram} and the accompanying text for some further
information on the effects of a non-zero initial strain rate.

\subsection{Discussion}
In the previous section, a series of necking simulations with different values of the
parameters $\epsilon_0$ and $D^\text{tot}$ was described. Trends seen in these
simulations are discussed in more depth in the text below.

\begin{figure}[p]
\begin{center}
      \includegraphics[width=\columnwidth,clip]{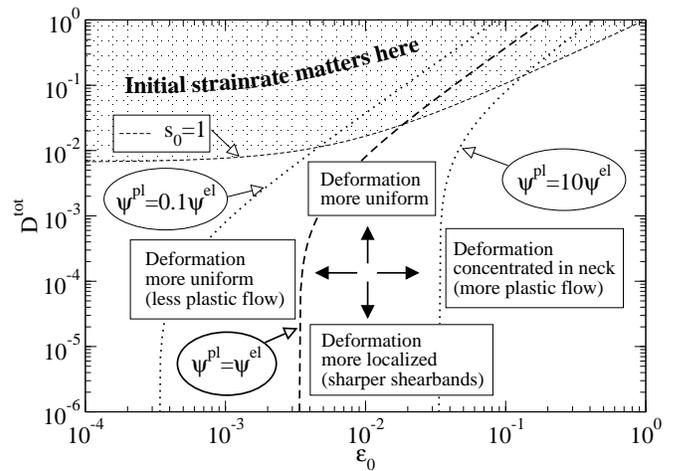}
      \caption{A ``phase
      diagram'' for the dynamics of constant strain rate necking simulations in the
      ($D^\text{tot}$, $\epsilon_0$) parameter space when $\Lambda\equiv 1$, $\mu=100$,
      and $\nu_2=0.5$. Simulations with a lower total strain rate had more pronounced
      shear bands. An increased $\epsilon_0$ resulted in more plastic flow, which
      allowed for sharper necks (in both time and space). The three curves running
      down the center indicate various values (at steady state) of the ratio
      of stored plastic to stored elastic energy, $\psi^\text{pl}/\psi^\text{el}$.
      If parameters were chosen from the
      shaded area at the top, the initial stress
      would be higher than the yield stress;
      thus, the strain rate would have to be ramped up from a lower value (perhaps
      from rest), and the details of how the strain rate was ramped up would affect
      the outcome of the simulation. The simulations in this paper deliberately
      avoided parameters from this region.}
      \label{fig:phase_diagram}
\end{center}
\end{figure}

Fig.~\ref{fig:phase_diagram} is a ``phase diagram'' of the
dynamics of the necking 
simulations, mapping out the behavior as a function of $D^\text{tot}$, the total
strain rate, and $\epsilon_0$, the strain due to flipped STZs (Spaepen uses a similar
diagram, which he calls a ``deformation map'', when he outlines the behavior of
amorphous metals in a graph of the temperature versus the stress
\cite{SPAEPEN_1977_AM_25_4_407, SPAEPEN_BOOK}).
It tries to capture the essence of the behavior described in the simulations above.

\begin{figure}
\begin{center}
      \includegraphics[width=\columnwidth,clip]{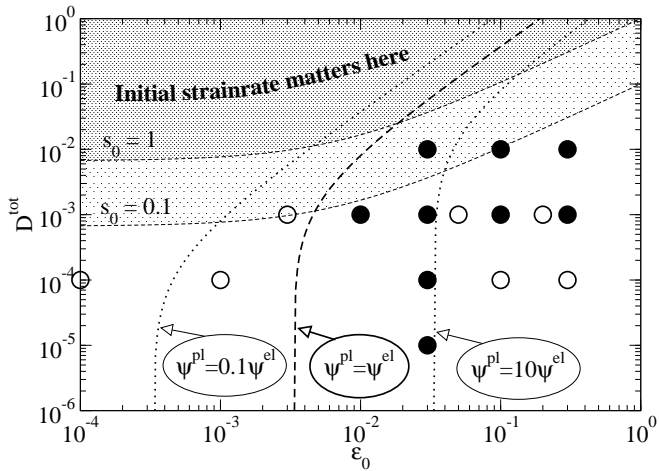}
      \caption
      {This is the same ``phase diagram'' as in Fig.~\ref{fig:phase_diagram},
      with simulations marked as open and filled circles. The latter are the
      simulations that were presented in Section~\ref{sec:simulations}. The shaded
      area added since Fig.~\ref{fig:phase_diagram} represents the parameter space
      where the initial stress would lie between $s_0=0.1$ and $s_0=1$. ($\mu=100$,
      $\nu_2=0.5$, $\Lambda\equiv 1$)}
      \label{fig:sims_Lambda1}
\end{center}
\end{figure}

Fig.~\ref{fig:sims_Lambda1} shows how the simulations discussed in the previous
section fit into the ``phase diagram'' of Fig.~\ref{fig:phase_diagram}.
Since there was more room in this graph, an additional curve and shading was added to
show what parameter values would cause the initial stress to lie between $s_0=0.1$ and
$s_0=1$, the latter being the yield stress.

Returning to Fig.~\ref{fig:phase_diagram}, the following is an attempt to explain the
trends seen there. First, the simulations with higher strain rates had
less pronounced shear bands. Since energy was supplied at a faster rate, the material
also had to dissipate this energy if it were to stay in steady state.  In the slow
strain rate simulations, it seems that the rate was slow enough that all the
dissipated energy could exit through the shear bands; in other words, only the shear
bands had stresses above the yield stress. In the simulations with higher strain rates
the input of energy per unit time was higher, resulting in energy dissipation
throughout the material. This forced the stress to exceed the yield stress
everywhere. Remember that if $\bar{s}>1$ then $\bar{\Delta}\to 1/\bar{s}$, while if
$\bar{s}<1$ then $\bar{\Delta}\to \bar{s}$. Since the plastic flow was proportional to
$\epsilon_0(s-\Delta)$ ($\Lambda_0=1$ in these simulations), the slowly strained
materials saw a large difference between the amount of plastic flow in the shear bands
and the rest of the material. The stresses would be above the yield stress in the
shear bands and below it elsewhere. Notice how the inhomogeneous geometry played a
vital role by breaking the symmetry, concentrating the stresses in certain areas, and
thus allowing the shear bands to form.

In contrast, the value of the strain rate had little effect on the simulations before
the stress reached the yield stress. As long as $\bar{\Delta}$ was low, $\Gamma$ was
negligible, and few STZs were created or annihilated. This implies that the
deformation before yield was almost reversible, and that the plastic strain was
proportional to $\bar{\Delta}$, the fraction of flipped STZs. Thus the material would
always reach the yield stress at the same total (that is, elastic and plastic) strain.

Second, when the parameter $\epsilon_0$ was given high values, the material would have
to be strained further before reaching yield stress. This parameter controlled the
amount of strain due to flipping STZs, or alternatively, how much of the energy was
absorbed in plastic deformations. By combining
Eqs.~\eqref{eq:stored_plastic} and \eqref{eq:stored_elastic}, one can see that a uniform
system in steady state with $s\approx 1$ has $\psi^\text{pl}/\psi^{el} \approx
\epsilon_0\,\mu(1+\nu_2)$\cite{LANCE-THESIS}; when this expression was less than
unity, most of the strain energy was stored as elastic strain. This would correspond
to $\epsilon_0<0.007$ in the above simulations, since $\mu=100$ and $\nu_2=0.5$. As
$\epsilon_0$ was increased, the STZs were responsible for more of the deformation, and
since the stress is only proportional to the elastic strain, the stress would grow
more slowly for large $\epsilon_0$. That explains why the materials with the smaller
$\epsilon_0$ yielded earlier.

Observing the post-yield dynamics, a large value of $\epsilon_0$ ultimately resulted
in sharper necks. After the stress had reached yield, the energy supplied by the work
at the grips would no longer be stored in the elastic or plastic deformations (which
corresponds to stress and flipped STZs, respectively). Rather, the energy was
dissipated through flipping newly created STZs while annihilating the same amount of
already-flipped ones. In a homogeneous system the material would just remain in this
steady state. In the necking simulations, though, the localized stresses emanating
from the indentations perturbed the system enough to change the flow from homogeneous
to inhomogeneous. Especially for higher values of $\epsilon_0$, the neck would narrow
rapidly, increasing the stress and plastic flow there until the material would snap in
two (in practice, the restricted geometry of the numerical grid would not actually
allow the material to split, but the neck would become extremely narrow and the
time-step would decrease to a value which was, for all practical purposes, zero). This
uncontrolled necking was mainly driven by stored energy being released through plastic
dissipation at the neck. Interestingly enough, it was the stored plastic energy that
supplied almost all of the energy; the stored elastic energy was almost negligible
when $\epsilon_0\gg 1/\mu(1+\nu_2)$.  An interesting question is: How much does the
quasi-linear approximation affect the dynamics? In a fully non-linear version of the
theory, the STZ transition rates would grow small for low stresses, thus preventing
the STZs from flipping back
\cite{FALK_1998_PRE_57_6_7192,LANGER_2003_PRE_68_6_061507}. Potentially, this could
prevent some of the release of the stored plastic energy from the areas further from
the neck.

A large value of $\epsilon_0$ meant that the material needed to be strained further
before the stress reached the yield stress since the plastic flow relaxed some of the
stresses. Although it took longer for the material to reach the yield stress, once it
did the neck developed more rapidly. This was because the increased plastic flow in
the neck allowed the ends to relax and release its stored plastic and elastic energy
faster. The relaxation of the stress at the ends added even more strain to the center,
contributing to the plastic flow at the neck. Finally, as the neck became thinner, the
stress would rise there and increase the already high plastic flow even
further. Thus the material would neck faster when $\epsilon_0$ was high, since that
would reinforce this feedback loop.

When $\epsilon_0\ll 1/\mu(1+\nu_2)$, most of the initial work done on the system was
stored as elastic energy. After reaching yield stress, the steady-state stress became
large (and $\bar{\Delta}=1/\bar{s}$ small) in order to create enough plastic
dissipation $\bar{D}^\text{pl}$ to
counter the steady input of energy supplied by the work done at the ends of the
material. There were some faint shear bands where the stresses showed slightly
elevated values, but the plastic dissipation was almost uniform throughout the
material when $\epsilon_0$ was small.


\section{Pre-Annealed Materials}
\label{sec:annealed}
The behavior of a pre-annealed metallic glass is more brittle than both a virgin
as-quenched sample and a material that has experienced plastic work. Experiments have
shown that metallic glasses that have been annealed below the glass transition
temperature seem more brittle, show more pronounced strain softening, and have
decreased plastic flow \cite{SPAEPEN_BOOK, SUH_2003_JNS_317_1-2_181,
HARMS_2003_JNS_317_1-2_200, FEDOROV_2001_TP_46_6_673, DEHEY_1998_AM_46_16_5873,
INOUE_1993_JNS_156__598, PAMPILLO_1978_MSE_33_2_275}. Microscopically, observations
showed structural relaxation in the form of more closely packed molecules,
corresponding to a macroscopic increase in density. Upon plastic deformation the
materials were seen to return to their as-quenched states, including a
decrease in the density and a lower packing fraction. Annealing for a longer time or
with a higher temperature gave more pronounced changes in the mechanical
properties. It is also worth noting that aging the materials over longer periods of
time at lower temperatures also made the metallic glasses more brittle
\cite{ARGON_1982_JPCS_43_10_945}.

The STZ theory captures the change in behavior due to annealing through $\Lambda$, the
relative density of STZs.  When a virgin material is initially quenched from a molten
state, the atoms have little time to organize into a closely packed configuration,
leaving a ``fluffy'' structure and a high density of STZs. Annealing the material
packs the atoms into a tighter configuration, making $\Lambda_0<1$. Likewise, if an
annealed material is subjected to plastic work, the value of $\Lambda$ would rise as
the tightly packed atoms are ``re-fluffed'', leaving more room for local
rearrangements.

The version of the STZ theory that is explored numerically in this paper has no
mechanism for thermal creep or relaxation, which is reflected in the fact that
$\dot{\Lambda}\geq 0$. Thus the model can only simulate \emph{pre}-annealed materials,
by setting the initial value of $\Lambda$ low \footnote{In order to incorporate the
effects of annealing and aging seen in real materials, there would need to be a
mechanism in Eq.~\eqref{eq:complete_Lambdadot} that could reduce the value of
$\Lambda$. There are currently efforts to incorporate thermal effects into the STZ
theory \cite{FALK_2004_PRE_70_1_011507, LANGER_2004_PRE_70_4_041502}.}. Consequently,
a material with $\Lambda_0<1$ and $\Lambda_0=1$ will from now on be referred to as
``pre-annealed'' and ``worked'', respectively, although the latter could also
represent a material in its virgin as-quenched state.

Initially during a constant
strain-rate experiment, a small $\Lambda$ would suppress plastic deformation since
there would be very few STZs to flip. As the stress
$\bar{s}\to 1$, most of the existing
STZs would be flipped, and new ones would have to be created causing $\Lambda$ to
grow. The lower $\epsilon_0$ was set, the higher the peak stress and steady-state
stress would be, and the faster $\Lambda$ would grow toward one.

The previous section assumed that $\Lambda\equiv 1$ throughout. The current section
describes simulations where pre-annealed materials were used during loading,
implemented by setting $\Lambda_0=0.01$. A spatially extended two-dimensional
simulation allowed $\Lambda$ to grow locally, thus making it possible for the material
to boost the plastic flow in areas where the stresses were high. As will be seen, this
contribution to the inhomogeneous deformation was particularly pronounced for high
values of $\epsilon_0$.

As in Section~\ref{sec:necking}, the simulations had a shear modulus of $\mu=100$, a
density of $\rho=1$, a Poisson ratio of $\nu_2=0.5$, and a numerical viscosity of
$\eta=0.02$. The size of the material was $4\times 4$, although this time the whole
material was simulated (that is, there were no symmetric boundaries). The material was
mapped onto a grid measuring $33\times 33$ nodes. The left and right boundaries
$X^0(y,t)$ and $X^1(y,t)$ (initially parallel to the $y$-axis) were both allowed to
deform. The lower boundary was held fixed at $y=0$, while the upper boundary $Y(t)$
was moved at a constant strain rate (see Section~\ref{sec:numerics} for further
details). The initial density of STZs was set below unity, to $\Lambda_0=0.01$. In
order to encourage inhomogeneous flow, the free boundaries were made jagged by
randomly perturbing the width with values up to $1\%$ of the total width (in other
words, after the perturbation $X^1(y,0)-X^0(y,0)\in[3.96,4.04]$). By setting the seed
for the random number generator to the same value for all the simulations, the initial
geometry would always be the same.

The simulations in this section illustrate the dynamics of pre-annealed materials in
three different ways. First, the distribution and average density of STZs are compared
to the dissipated energy and the work done on the systems for different values of
$\epsilon_0$. Second, the dynamics of pre-annealed and worked materials are compared.
And third, the following question is addressed: Would the simulations
with higher strain rates experience more pronounced shear bands if they were stopped
and held at a fixed strain for the same amount of time it would take a slower
simulation to reach that same strain?

\subsection{Variation in $\epsilon_0$}
\begin{figure*}
  \begin{tabular}{c}
      \fbox{\includegraphics[height=0.29\textheight,clip]{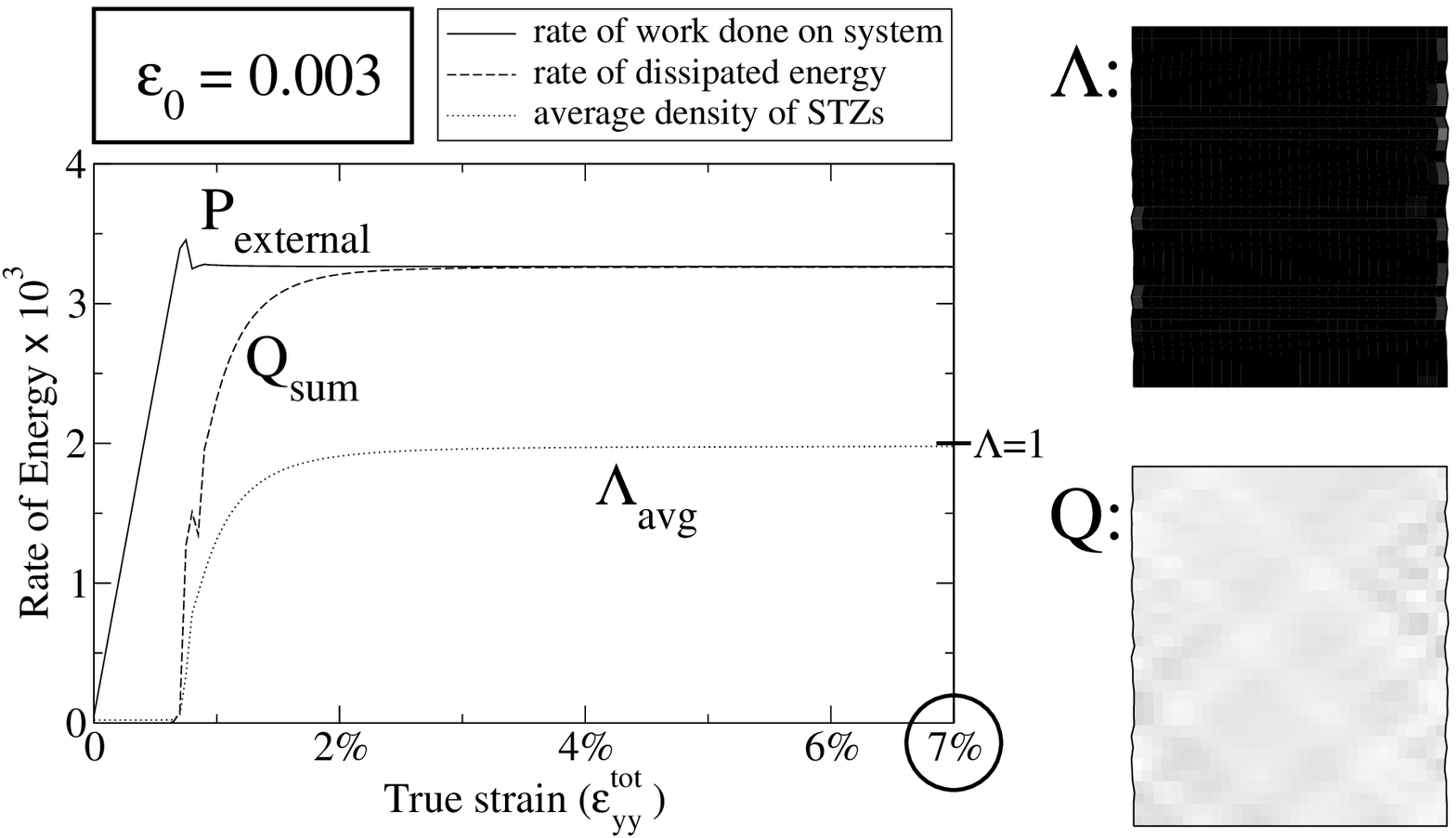}}\\
      \fbox{\includegraphics[height=0.29\textheight,clip]{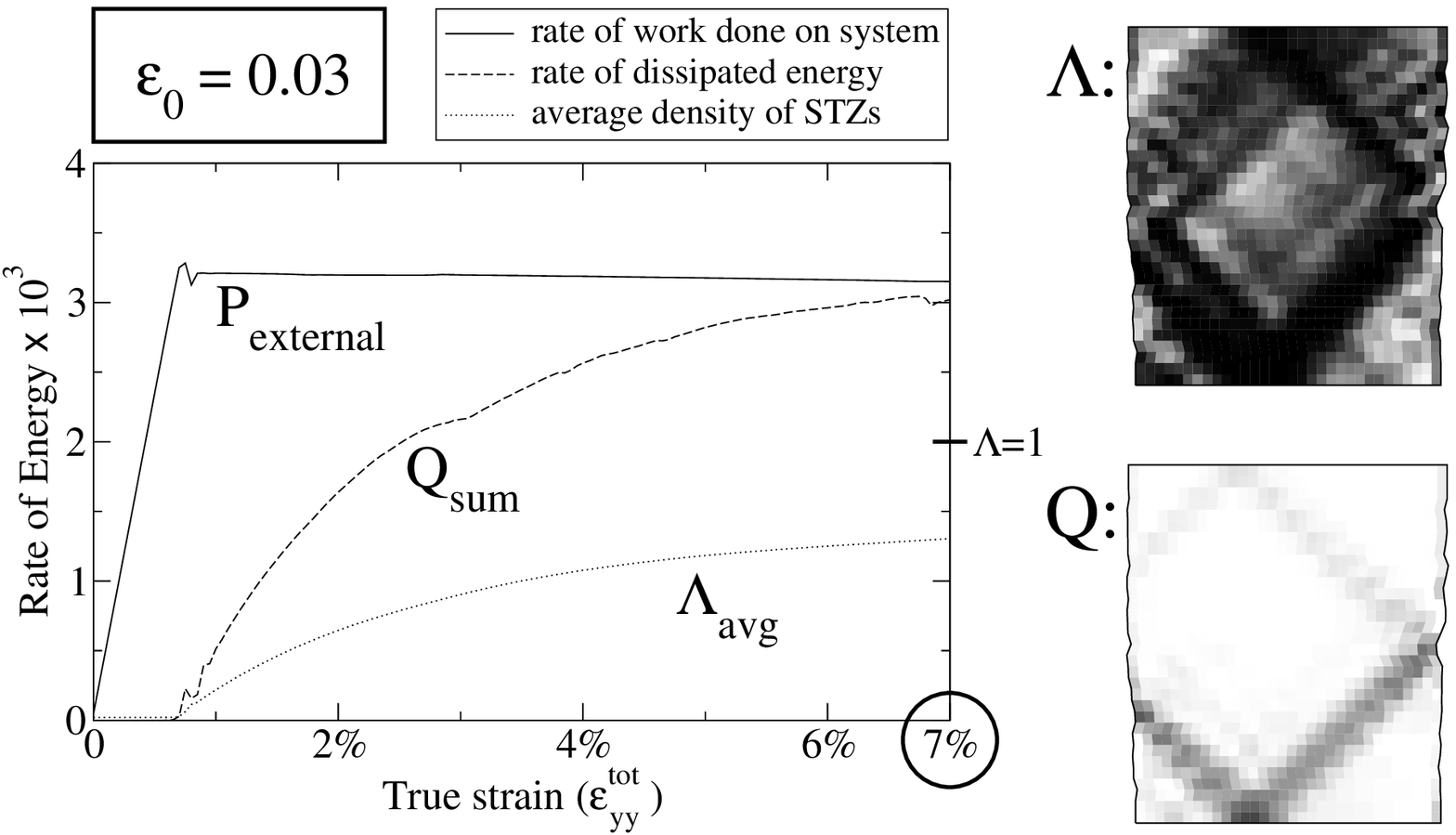}}\\
      \fbox{\includegraphics[height=0.29\textheight,clip]{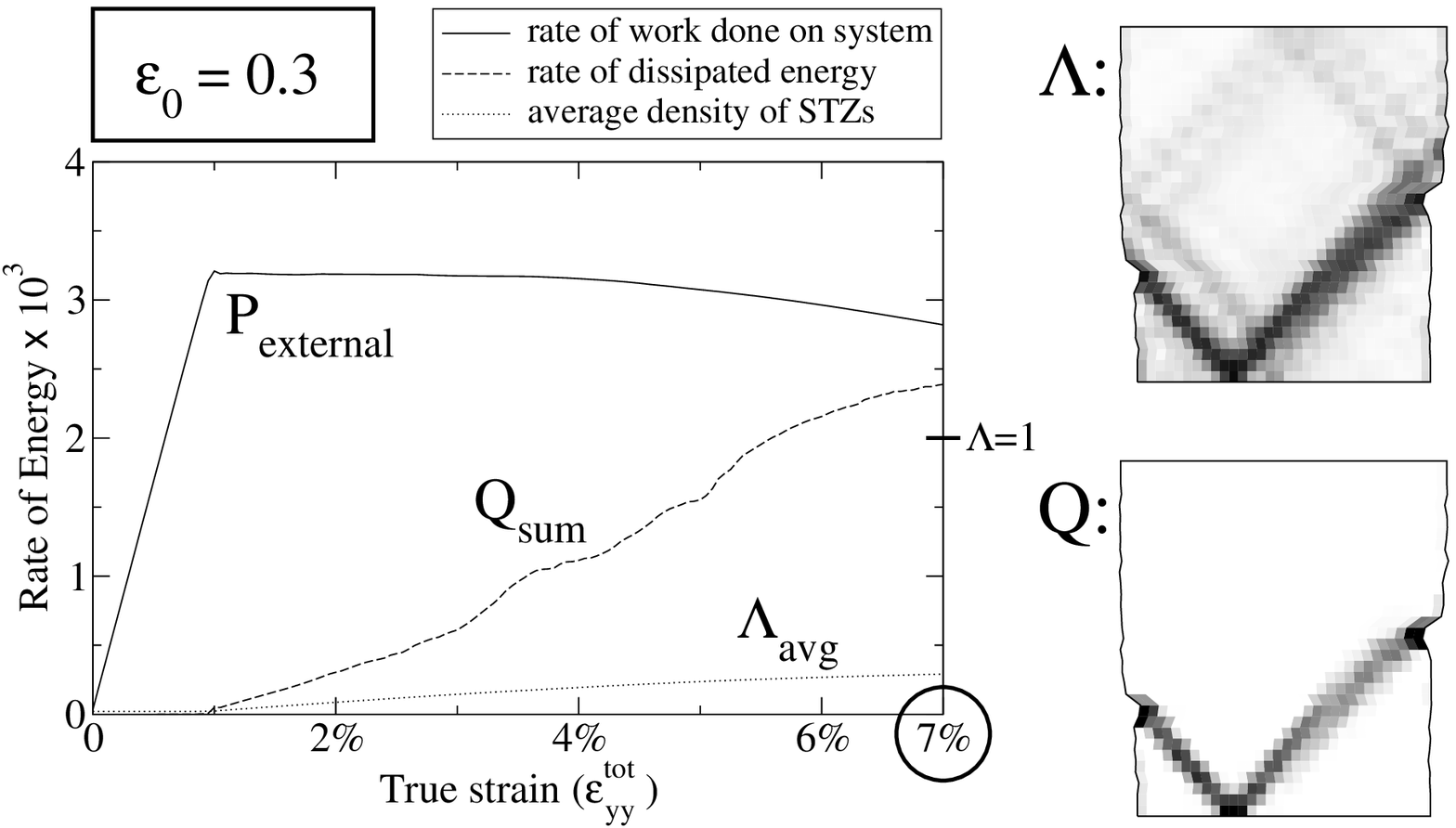}}
  \end{tabular}
  \caption{Three pre-annealed simulations with $\Lambda_0=0.01$ were run for
      $\epsilon_0=0.003$, $\epsilon_0=0.03$, and $\epsilon_0=0.3$, all with a strain
      rate of $D^\text{tot}=10^{-4}$. The graphs on the left show $P_\text{external}$,
      the rate of work done at the grips, $Q_\text{sum}$, the rate of dissipated
      plastic energy summed over the whole material, and $\Lambda_\text{avg}$, the
      average density of STZs. The left vertical axis shows the range for
      $P_\text{external}$ and $Q_\text{sum}$ (the interval is $[0,0.004]$), while the
      right vertical axis displays the scale for $\Lambda_\text{avg}$. The density
      plots on the right show $\Lambda$ and $Q$ at $7\%$ strain.}
  \label{fig:energy_and_Lambda_various_eps0}
\end{figure*}

Fig.~\ref{fig:energy_and_Lambda_various_eps0} compares the average density of STZs
\begin{equation}
\Lambda_\text{avg} \equiv \frac{1}{A}\int_A\Lambda\,dA
\end{equation}
(where $A$ is the area of the sample) to the plastic dissipation of energy
\begin{equation}
Q_\text{sum} \equiv \int_AQ\,dA
\end{equation}
for three different values of $\epsilon_0$.
All three simulations were strained at a rate of
$D^\text{tot}=10^{-4}$. Each of the three graphs on the left displays three curves, of
which the solid represents the rate of work done on the system at the grip,
\begin{equation}
P_\text{external} = F\,v_y = (X^1-X^0)\sigma_{yy}v_y\,.
\end{equation}
$F$ is the force applied to the end of the material, and $\sigma_{yy}$, $v_y$, $X^0$,
and $X^1$ were all evaluated at $y=Y(t)$. Since the upper boundary was kept flat and
moved perpendicular to 
its surface, only the $y$-component of the stress was needed. $P_\text{external}$ is
thus the rate at which energy flowed into the system.

The middle curve in all three graphs shows $Q_\text{sum}$, the rate of energy
dissipated due to plastic deformation summed over the whole material. The axis on the
left side displays the values for both this curve and that of $P_\text{external}$,
although they were multiplied by $10^3$ to reduce clutter (the interval is really
$[0,0.004]$). The axis on the right sets the scale for the third curve,
$\Lambda_\text{avg}$, which represents the density of STZs averaged over the whole
sample.

The two density plots to the right of each of the graphs show the final distributions
of $\Lambda$ and $Q$ when the systems had reached $7\%$ strain. In the density plots
for $\Lambda$, the shading was scaled so that the interval $[0,1]$ went from white to
black. With $Q$, the gray scale was mapped onto the interval $[0,0.003]$, with white
again representing zero. In the latter density plots there were occasional points that
exceeded $0.003$ (these were just painted black as well), but choosing a larger
interval would have erased most of the structure seen in the pictures.

In the simulation where $\epsilon_0=0.003$, $\Lambda_\text{avg}$ quickly rose to one
while $Q_\text{sum}$
grew equal to $P_\text{external}$ once the stress reached yield. The latter meant that
all the energy provided at the grips was dissipated through plastic deformation, and
none was stored, after the material reached about $2\%$ strain. In the
$\epsilon_0=0.3$ simulations, $\Lambda_\text{avg}$ never even reached a value of
$0.15$, and the plastic rate of dissipation grew a lot slower. In fact, looking at the
density plot shows that the density of STZs only saturated along one band, and the
plot for $Q$ shows that this was where most of the plastic
dissipation took place as well. The fact that $\Lambda$ so quickly reached its
equilibrium value everywhere in the material for small $\epsilon_0$ meant that the
initial value of $\Lambda_0=0.01$ had little effect in this case.

Although the density plots of $Q$ and $\Lambda$ both were snapshots, the latter in
some ways gave a cumulative look at what had happened since the start, since $\Lambda$
could not decay (this is not true in other versions of the STZ theory that incorporate
thermal relaxation \cite{FALK_2004_PRE_70_1_011507, LANGER_2004_PRE_70_4_041502}). In
contrast, the density plot of $Q$ is more appropriate when examining the instantaneous
dynamics, since it highlights the current rate of plastic deformation in the different
areas of the material.

In the simulation with $\epsilon_0=0.3$, the rate at which energy flowed into the
system, $P_\text{external}$, started decaying towards the end. This is because the
stress at the grip began to drop due to the increased plastic flow
along the shear band. The high value of $\epsilon_0$ allowed the material to deform
enough in the small area of this band to account for all the strain imposed by the
grip, thus inducing inhomogeneous flow and perhaps cause the material to break at a
later time.


\subsection{Comparing Pre-Annealed and Worked Materials}
\begin{figure}
\begin{center}
      \includegraphics[width=\columnwidth,clip]{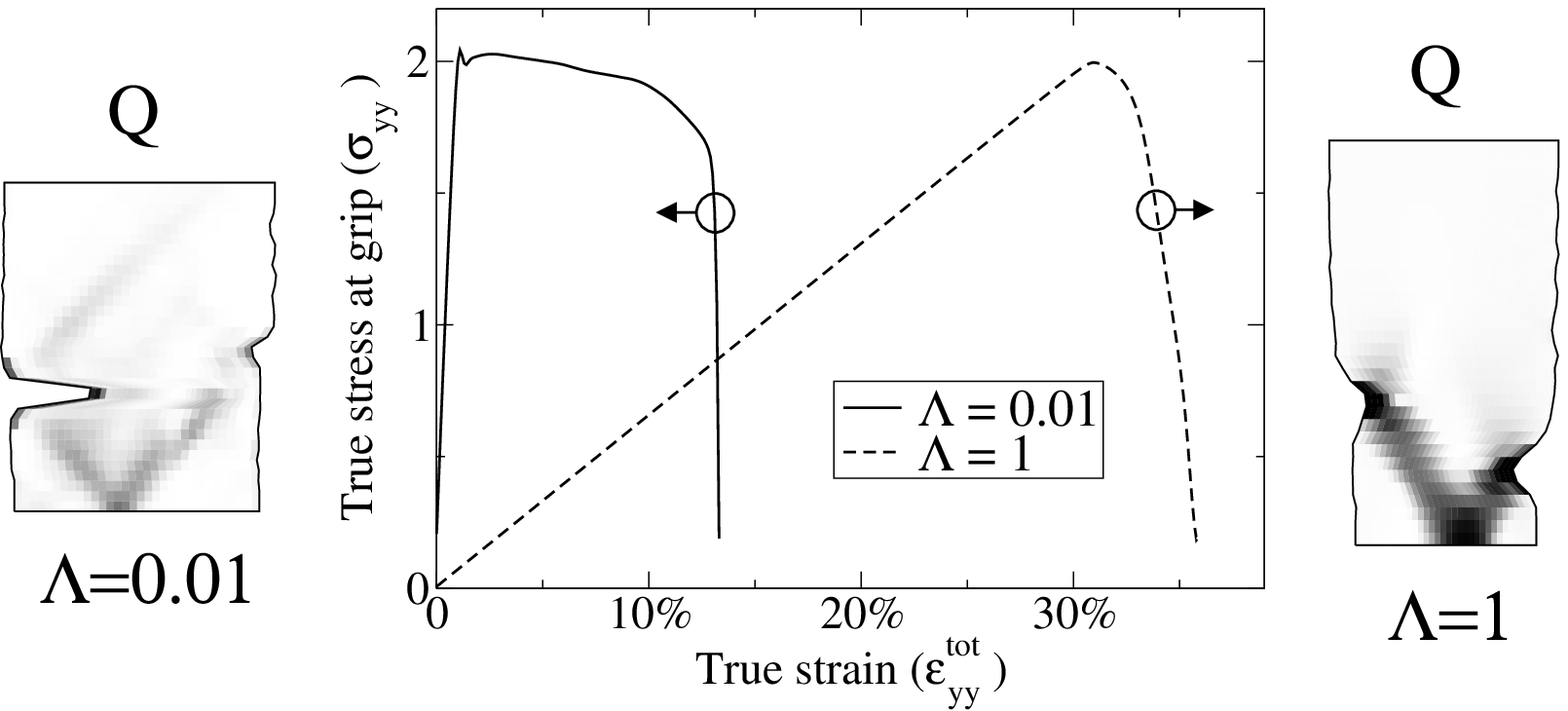}
      \caption
      {The true stress-strain curves for two materials, one with
      $\Lambda_0=0.01$ and one with $\Lambda_0=1$. The density plots show the
      plastic rate of dissipation at some arbitrary time after yield, with the stress
      at the grips being the same in the left and right
      snapshots. ($D^\text{tot}=10^{-3}$, $\epsilon_0=0.3$)}
      \label{fig:compare_Lambda_break_dtot0.001}
\end{center}
\end{figure}

Fig.~\ref{fig:compare_Lambda_break_dtot0.001} compares a pre-annealed and a worked
material, having initial conditions of $\Lambda_0=0.01$ and $\Lambda_0=1$,
respectively.  The two curves in the center show the true stress-strain curves of
these two materials as they were strained beyond yield and toward failure (the
materials would eventually grow extremely thin somewhere in the center). The density
plots of $Q$ on either side were arbitrarily chosen some time after the materials had
yielded, with the stress at the grips being the same in the two snapshots. Although
the two density plots look different, their initial geometries were identical to each
other and to all the other simulations in this section.

The only comparison shown here is that of the simulations with
$\epsilon_0=0.3$. Although no less important, the pre-annealed materials with a lower
$\epsilon_0$ would have $\Lambda$ rise to one everywhere so quickly that the results
were essentially the same as for the worked materials; this effect was seen in
Fig.~\ref{fig:energy_and_Lambda_various_eps0}.

A high value of $\epsilon_0$ allowed the stresses to grow more slowly, permitting
some areas to reach $\bar{s}>1$ while others remained at $\bar{s}<1$ as the material
approached yield. $\Lambda$ would only grow in the resulting shear bands, thus making
its average value grow slowly. A slowly growing stress (with respect to time) would also
occur for small strain rates, resulting in a similar shear-banding effect.

Comparing the stress-strain curves in Fig.~\ref{fig:compare_Lambda_break_dtot0.001},
the pre-annealed material plat\-eaued before breaking. It almost seemed like it was riding
along an unstable equilibrium before it increased the density of STZs in one location
and then deformed and broke there. In comparison, the worked material would deform
substantially more before reaching yield stress, and then ``ooze'' apart (rather than
``break'').

It is tempting, although perhaps somewhat speculative, to compare the necking on the
left side of Fig.~\ref{fig:compare_Lambda_break_dtot0.001} to fracture. Some of the
simulations, including this one, had extremely concentrated stresses and narrow necks,
and the deformations behaved in many ways similar to brittle cracks. The simulations
have shown that there is an interplay between the STZs, the geometry, and stress
concentrations. Naturally, the geometry of the numerical grid in these simulations was
quite restricted, the
single-valued boundaries
forcing the ``crack'' to run horizontally rather than following the $45^\circ$ shear
bands as seen in experiments \cite{PAMPILLO_1974_JMS_9_5_718,
PAMPILLO_1975_JMS_10_7_1194}. Also, the resolution was too low to make any good
quantitative arguments. Nevertheless, this behavior suggests that the STZ description
might be capturing some of the dynamics that is present in fracture (for more
information on fracture in brittle amorphous materials, see the review by Fineberg and
Marder \cite{FINEBERG_1999_PRSPL_313_1-2_2}).

\subsection{Strain-Rate Dependent Localization}
When comparing the results of the simulations from both the current and previous
sections, the 
shear bands at a given strain were more pronounced for smaller strain rates. Comparing
two identical materials that were strained to $7\%$ at rates
of $D^\text{tot}=10^{-4}$ and $D^\text{tot}=10^{-3}$, the former would not only be
stretched ten times slower, it would also have ten times as long to relax and
deform. Could it be that it was not the difference in speed, but the difference in
relaxation time that allowed the shear bands to form in the slow case?

\begin{figure}
\begin{center}
      \includegraphics[width=\columnwidth,clip]{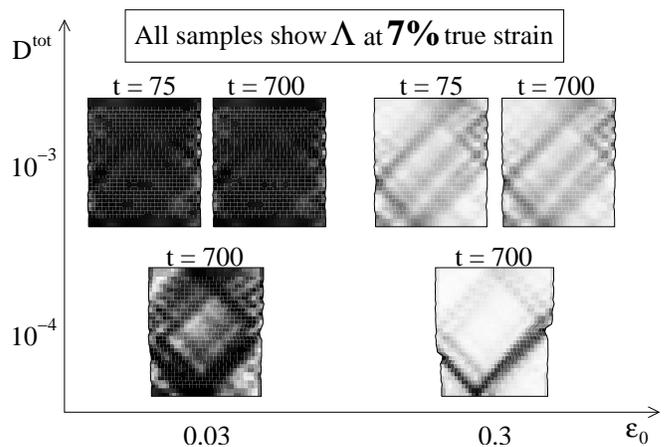}
      \caption
      {Density plots of $\Lambda$ at $7\%$ strain, both for $\epsilon_0=0.03$
      and $\epsilon_0=0.3$. The systems were strained at rates of
      $D^\text{tot}=10^{-4}$ and $D^\text{tot}=10^{-3}$, and in the latter case the
      grip was held fixed at $7\%$ for the same amount of time that it took the slower
      strain rate to reach this strain. Despite the extra time, no pronounced
      shear bands developed; in fact, once the grip was stopped, the material hardly
      changed at all.}
      \label{fig:stopped_Lambda}
\end{center}
\end{figure}

Fig.~\ref{fig:stopped_Lambda} shows density plots of $\Lambda$ at $7\%$ strain where
the materials had been strained at the rates mentioned above, with both
$\epsilon_0=0.03$ and $\epsilon_0=0.3$.
The system that had been strained at the
slower rate, $D^\text{tot}=10^{-4}$, reached $7\%$ at time $t=700$, while the faster
system with a rate of $D^\text{tot}=10^{-3}$ was stopped around $t=70$ (the snapshot
was taken at $t=75$ to make sure the material had come to a complete stop, reaching
its full $7\%$; in order to stop the material at this exact strain, it was necessary
to start slowing it down already at $t=65$). After stopping the fast system, it was
held fixed until $t=700$, the same amount of time it took the slow system to reach
$7\%$ strain.

It turned out that there was practically no change in the quickly-strained material
after the grips had come to a halt (this was the case both for $\epsilon_0=0.03$ and
$\epsilon_0=0.3$, although the former ended up with a higher
$\Lambda_\text{avg}$). This means that all the deformation took place instantly, not
allowing any of the stored energy to escape later.

This does not preclude a situation where a more irregular geometry might induce a
stress concentration, allowing both stored plastic and elastic energy to be released,
perhaps even driving a necking instability; this kind of behavior was seen in the
previous section. It does mean, though, that the shear bands were created without the
need of instability mechanisms. In fact, the bands of STZs were more pronounced when
energy flowed into the system at a slower rate.

The previous section speculated that
the STZ theory could perhaps contribute to explain fracture dynamics. The
current section seems to contradict this somewhat, at least at first glance. From
experience, a material strained at a higher rate seems more brittle, implying that it
should have sharper stress concentrations, and more localized shear bands. How could
such behavior be compatible with the results shown above?

It was briefly mentioned earlier that the thermal relaxation is being incorporated
into some versions of the STZ theory \cite{FALK_2004_PRE_70_1_011507,
LANGER_2004_PRE_70_4_041502}. With that mechanism included, $\Lambda$ can decrease as
STZs are annihilated over time, and this might suppress the creation of localized
bands of STZs as the strain rate drops. The model used in this paper did not allow
$\Lambda$ to decrease, which might be interpreted as running the simulations close to
zero temperature.


\section{Conclusion}
\label{sec:conclusion}
The goal of this paper has been to explore the STZ theory in a spatially extended
geometry. In addition to showing that a two-dimensional implementation is
capable of describing shear localization, 
a wide range of parameter values were used to expose the different types of behavior
inherent in the model. The geometry, energy flow, and internal state of the material
all contributed to the rheology through effects such as jamming and plastic flow,
annealing, strain softening, necking, and shear banding.

Amorphous metals have two modes of deformation: Homogeneous and inhomogeneous flow
\cite{SPAEPEN_1977_AM_25_4_407, SPAEPEN_BOOK, NIEH_2002_I_10_11-12_1177}. Homogeneous
flow is typical for low stresses and high temperatures, and under uniaxial tension a
material sample will deform uniformly throughout the specimen. In inhomogeneous flow,
the deformation is usually localized in narrow shear bands that run at a $45^\circ$
angle with respect to the tensile axis \cite{PAMPILLO_1974_JMS_9_5_718,
PAMPILLO_1975_JMS_10_7_1194} (some studies show that the angle of the shear bands
deviate under large isotropic pressures
\cite{LEWANDOWSKI_2002_PMACMSDMP_82_17-18_3427, LOWHAPHANDU_1999_SM_41_1_19}). The
cross-section of the material decreases as slip, and eventually fracture, occurs along
these shear bands \cite{SPAEPEN_BOOK, PAMPILLO_1974_JMS_9_5_718,
CHOU_1975_AM_23_5_609}. At low temperatures, the shear-localization instability sets
in right after yield, making the material behave in a seemingly brittle manner; there
is no hardening due to strain in metallic glasses, although physical aging decreases
the plastic response \cite{ARGON_1982_JPCS_43_10_945}.

Section~\ref{sec:necking} looked at how the strain rate, the straining-capability of
STZs ($\epsilon_0$), and the geometry affected flow and deformation during necking. It
was especially striking how sharp shear bands and narrow necks were formed at low
strain rates and high $\epsilon_0$, respectively.  The former of these two trends
appeared to contradict experimental evidence: An amorphous metal displays increased
brittle behavior as it is strained at a higher rate \cite{LU_2003_AM_51_12_3429},
while the opposite seemed true for the simulated materials. The discrepancy probably
stems from the lack of temperature in the current STZ model. There was no mechanism
once the simulation had started, apart from plastic deformation, to annihilate
existing STZs. In the experiments, the non-zero temperature allowed the molecules in
the most strained areas to relax if given enough time. This resulted in more
homogeneous flow as the material was strained at lower rates. Efforts have been made
to incorporate thermal relaxation into the model \cite{FALK_2004_PRE_70_1_011507,
LANGER_2004_PRE_70_4_041502}.

Section~\ref{sec:annealed} considered pre-annealed amorphous solids, starting with a
lower initial relative density of STZs $\Lambda_0$. Experiments have shown that
amorphous metals become more brittle when annealed, even though no crystallization is
detected \cite{SUH_2003_JNS_317_1-2_181, HARMS_2003_JNS_317_1-2_200,
FEDOROV_2001_TP_46_6_673, DEHEY_1998_AM_46_16_5873, INOUE_1993_JNS_156__598,
PAMPILLO_1978_MSE_33_2_275}. Rather, the embrittlement correlates with structural
relaxation, leaving the more closely packed molecules less room to move.  In the
simulations, the lower $\Lambda_0$ enhanced the localization of the strain, especially
for higher values of $\epsilon_0$ when most of the stored energy was in the form of
flipped STZs (rather than elastic strain).  Compared to either virgin as-quenched or
worked samples, the pre-annealed materials behaved in a more brittle manner with
narrower shear bands and something resembling a cleavage fracture. The latter
observation should be approached with caution since the implementation was not
designed to handle such extreme deformations; the numerical grid forced the ``crack''
to run horizontally through the material, while a diagonal path running at $45^\circ$
would seem more natural \footnote{Rather than tracking the boundaries by mapping the
coordinate system onto a unit square, the phase-field
\cite{EASTGATE_2002_PRE_65_3_036117} and level-set \cite{SETHIAN,OSHER} methods were
briefly considered; although computationally more expensive, they might be
necessary in more complicated and extreme geometries.}. Nevertheless, the similarities
between the dynamics of the STZ theory at low $\Lambda_0$ and empirically annealed
solids were strong enough to warrant further investigation.

As mentioned, an increased value of $\epsilon_0$ would result in more of the applied
work being stored as flipped STZs rather than elastic strain. This was particularly
apparent for the virgin materials ($\Lambda_0=1$), where an increased $\epsilon_0$
meant that the material would reach yield stress at a much larger total strain. On the
one hand, it is possible that the large values of $\epsilon_0$, which were needed to
produce interesting dynamics in the simulations, exaggerated the strain caused by
flipped STZs. On the other hand, $\Lambda_0\ll 1$ would restrict the plastic strain (and
thus the stored plastic energy) to narrow shear bands, causing a mostly elastic
behavior before yield and more brittle dynamics at failure. It is possible that the
sharp bands of inhomogeneous flow in experimentally deforming amorphous solids
are due to an internal structure corresponding to low initial values
of $\Lambda$; unless the materials were quenched extremely rapidly and to very low
temperatures, chances are that some structural relaxation would occur.
Incorporating the earlier mentioned thermal relaxation into the STZ model could help
lower the value of $\Lambda$ throughout the simulation and yield better agreement
with the experimental results.

This work was supported primarily by the Research Council of Norway. It was supported
in part by the National Science Foundation Materials Research Science and Engineering
Centers (NSF MRSEC) program (DMR96-32716), due to Jim Langer's gracious and
surprising decision to assist me financially when my main grant ended. This work made
use of the computing facility of the Cornell Center for Materials Research (CCMR) with
support from the NSF MRSEC program (DMR0079992), and computing facilities at UC Santa
Barbara sponsored by a grant from the Keck Foundation for Interdisciplinary Research
in Seismology and Materials Science. Additional support came from the NSF-KDI program
(DMR-9873214) and the U.S. Department of Energy (DE-FG03-99ER45762). I would like to
thank Jim Langer for his advice and support, as well as Leonid Pechenik, Craig
Maloney, Anthony Foglia, and Michael Falk for stimulating discussions.


\bibliography{non_jrt_refs,jrt_refs}

\end{document}